\documentclass[noshowpacs,twocolumn,aps]{revtex4}

\usepackage{graphicx} 
\usepackage{dcolumn}  
\usepackage{bm}       
\usepackage{amssymb}  
\usepackage{amsmath}  
\usepackage{setspace}
\usepackage{latexsym}
\usepackage{calc}
\usepackage{shadow}
\usepackage{epsfig}
\usepackage{palatino}
\usepackage[usenames]{color} 	  

\input epsf
\input rotate

\def\bea{\begin{eqnarray}}
\def\eea{\end{eqnarray}}
\def\ben{\begin{equation}}
\def\een{\end{equation}}
\def\benu{\begin{enumerate}}
\def\enu{\end{enumerate}}

\def\n{n}

\def\sss{\scriptscriptstyle\rm}

\def\1var{(\bx_1...\bx\N)}

\def\half{\frac{1}{2}}

\def\br{{\bf r}}

\def\bx{{\br t}}

\def\bj{{\bf j}}

\def\x{_{\sss X}}
\def\c{_{\sss C}}
\def\s{_{\sss S}}
\def\xc{_{\sss XC}}

\def\Hxc{_{\sss HXC}}

\def\N{_{\sss N}}
\def\H{_{\sss H}}

\def\ext{_{\rm ext}}

\def\ee{_{\rm ee}}

\def\sph_int{ {\int d^3 r}}

\def\bei{\begin{itemize}}
\def\eei{\end{itemize}}
\def\nn{\nonumber}

\begin{document}

\title{Propagation of Initially Excited States in Time-Dependent Density Functional Theory}

\author{Peter Elliott}
\affiliation{Department of Physics and Astronomy, Hunter College and the City University of New York, 695 Park Avenue, New York, New York 10065, USA}
\author{Neepa T. Maitra}
\affiliation{Department of Physics and Astronomy, Hunter College and the City University of New York, 695 Park Avenue, New York, New York 10065, USA}


\begin{abstract}
Many recent applications of time-dependent density functional theory
begin in an initially excited state, and propagate it using an
adiabatic approximation for the exchange-correlation potential. This
however inserts the excited-state density into a ground-state
approximation.  By studying a series of model calculations, we
highlight the relevance of initial-state dependence of the exact
functional when starting in an excited state, and explore the errors
inherent in the adiabatic approximation that neglect this dependence.
\end{abstract}

\maketitle

\section{Introduction}
\label{sec:Intro}
Time-dependent density functional theory (TDDFT) is an exact
reformulation of the time-dependent quantum mechanics of many-electron
systems~\cite{RG84,newTDDFTbook} that operates by mapping a system of
interacting electrons into one of fictitious non-interacting fermions, the Kohn-Sham (KS) system,
 reproducing the same time-dependent density of the true system. By
the theorems of TDDFT, all properties of the true interacting system
may be obtained from the KS system, in principle. In practice,
approximations are needed for the exchange-correlation effects: in
particular, for the exchange-correlation (xc) potential,
$v\xc[n;\Psi(0),\Phi(0)](\br,t)$, a crucial term in the
single-particle KS equations determining the evolution of the
ficitious non-interacting fermions. This functionally depends on the one-body density, $n(\br,t)$, the initial many-body state of the interacting system $\Psi(0)$, and
the initial state of the KS system $\Phi(0)$. 

From the inception of the theory, subtleties were noted in the
functional-dependences, which make good approximations for the
xc potential more challenging to derive than for its
counterpart in the (much older) ground-state density functional
theory~\cite{HK64,KS65,K99b}. One of these is memory: the dependence
of the potential at time $t$ on the density at earlier times, and on
the initial-states $\Psi(0)$ and $\Phi(0)$. Efforts to include
memory-dependence in
functionals~\cite{GK85,Db94,VK96,DBG97,VUC97,TP03,KB04,WU08,T12}, are not widely used, and have all focussed on the dependence on the history of the density, neglecting the dependence on the initial states.
 The vast majority of applications have been in linear response from
a non-degenerate ground-state, and there this initial-state dependence (ISD) is redundant: by the Hohenberg-Kohn
theorem, a non-degenerate ground-state is a functional of its own density.
In fact, the simple adiabatic approximation, where the instantaneous density is input into a ground-state approximation,
neglecting any memory-dependence, has shown
remarkable success for a great range of excitations and spectra. All commonly available codes with TDDFT capabilities are written assuming the adiabatic approximation. 
Over the years it has been understood that certain
types of excitations cannot be captured by the adiabatic approximation, e.g. double
excitations~\cite{MZCB04}, excitonic series in optical response of
semiconductors~\cite{GORT07}, and there is both on-going research in developing
frequency-dependent kernels to deal with this, as well as
understanding when not to trust the calculated adiabatic TDDFT
spectra.

With the results and understanding of recent years lending a level of
comfort with calculations of spectra and response, TDDFT has now
entered a more mature stage, and with that, comes more adventurous
applications. In particular, for electron dynamics in real time,
evolving under strong laser fields, or coupled electron-nuclear
dynamics following photo-excitation (e.g. Refs.~\cite{FB11,TTRF08,GROT09}).
The role of memory is less well-understood in these applications;
studies on model systems have shown that sometimes memory-dependence
is essential~\cite{AV99,HMB02,MB02,U06,FHTR11}, other times it is not
important at all~\cite{TGK08,B09}.  Moreover a new element enters: the
initial-state dependence that was conveniently and correctly brushed
aside in the linear response regime, now raises its head. In
applications such as modeling solar cell processes, the initial
photo-excitation of the electronic system is not dynamically modeled;
instead the dynamics begins with the electronic system assumed to be
in an excited state. The initial state is not the ground-state, yet no
truly initial-state dependent functionals are available today, hence
we ask how large are the errors in such calculations? Knowing that the
true interacting system begins in a certain excited state, is there an
optimal choice for the initial KS state when propagating with an
adiabatic approximation? How large are the errors due to ISD compared
to those due to history-dependence?

In this paper, we begin to answer these questions by considering a series of model
calculations of two-electron systems. We start by reviewing the
underlying theorems of TDDFT and the subtleties of ISD, even in
situations where we may not expect it. We show how ISD leads to a non-zero
xc potential even for non-interacting electrons.

Then, we move to electron dynamics in
the model soft-Coulomb helium atom, performing adiabatic TDDFT
calculations with different initial-states and comparing with exact dynamics. Here we must dissect other
sources of error in usual TDDFT calculations, such as using an initial KS excited state whose density does not equal the true excited state density,
in order to
properly ascribe the influence of ISD. Finally we make the first step towards investigating how ISD
can affect an area currently of much interest, namely coupled
electron-ion dynamics, by performing Ehrenfest calculations for a model
LiH system.  We work in atomic units throughout ($e^2 = m_e = \hbar =
1$).

\section{Initial-State Dependence in TDDFT}
\label{sec:Back}
In TDDFT, one evolves a set of single-particle orbitals \{$\phi_j(\br,t)$\} with a one-body KS potential:
\ben
i\frac{\partial}{\partial t}\phi_j(\br,t) = \left( -\half\nabla^2+ v\s(\br,t)\right)\phi_j(\br,t)
\een
\ben
v\s(\br,t) = v\ext(\br,t) + v\xc[n](\br,t) + v\xc[n;\Psi_0,\Phi_0](\br,t)
\label{eq:vks}
\een
where  $v\H(\br,t) = \int n(\br',t)/\vert \br - \br'\vert d^3r'$ is
the usual Hartree potential  and $v\xc(\br,t)$ is the xc potential which
depends on the entire history of the density, the interacting initial
state $\Psi_0$, and the initial KS wavefunction $\Phi_0$.

The origin of the initial-state dependence in the functionals is that
the one-to-one Runge-Gross mapping~\cite{RG84} between time-dependent
densities and potentials holds for a fixed initial-state. Thus, the
external potential $v\ext$ functionally depends on the density and the
true initial state $\Psi(0)$, and the KS potential $v\s$ depends on
the density and the KS initial state $\Phi(0)$. These functional
dependences are not directly relevant themselves in a practical
calculation, because there only the xc potential
$v\xc$ needs to be approximated. However, they lend their dependences
to $v\xc$ via Eq.~(\ref{eq:vks}), which must therefore depend on both
initial states and the density. 

There are two known situations in which there is no initial-state
dependence.  When the initial state is a ground-state, then, by the
Hohenberg-Kohn theorem of ground-state DFT, it is itself a functional
of its own density. ISD is redundant, as the information about the
initial state is contained in the initial density. (See also
Sec.~\ref{sec:gsisd}). The other situation is for one-electron
systems: starting in {\it any} initial state, there is only one
potential that can yield a given density-evolution~\cite{MB01}. In all
other situations, it is assumed that ISD cannot be subsumed into a
density-dependence; explicit demonstrations for two electrons can be
found in Refs.~\cite{MB01,MB02}. Ref.~\cite{MBW02} derived an exact
condition relating the dependence on the history of the density to
ISD. This condition is likely violated by any history-dependent
functional approximation that has no ISD.

Technically, one may choose any initial KS wavefunction that has the
same $n$ and $\dot{n}$ as the true initial state~\cite{L99}.  Usually
a single Slater-determinant of $N$ spin-orbitals $\phi_i$ is selected, with the required property that 
\bea
\label{eq:n0}
\sum_{i=1}^{N}\vert \phi_i(\br,0)\vert^2 &=& n(\br,0)\; {\rm and}\\
-\nabla\cdot {\rm Im} \sum_{i=1}^N \phi_i^*(\br,0)\nabla\phi_i(\br,0) &=&\dot{n}(\br,0)
\label{eq:ndot0}
\eea
where

\ben
\n(\br,0) = N \sum_\sigma\int{dx_2}\cdots\int{dx_N} \vert \Psi(x, x_2\cdots x_N,0)\vert^2
\een

\bea
\dot{\n}(\br,0) =& -N \nabla\cdot {\rm Im}  \ &\left\{ \sum_\sigma  \int dx_2\cdots\int dx_N   \Psi^*(x,x_2…x_N,0) \right.\nn \\
&&\left. \nabla\Psi(x,x_2,…,x_N,0) \vphantom{\sum_\sigma} \right\}
\eea
where $x=(\br,\sigma)$ and $\int dx=\sum_\sigma\int d^3x$ .

In this paper we will consider different choices of $\Phi(0)$ for
two-electron systems that begin in the first excited singlet state of the true
system, denoted $\Psi^*$. Specifically we will, at various points, investigate the following three forms:
\newline
(a) An excited KS singlet state  with two occupied orbitals $\phi_0$ and $\phi_1$, whose spatial part has the form
\ben
\label{dsd}
\Phi^*(\br_1,\br_2) = \frac{1}{\sqrt{2}}\left( \phi_0(\br_1)\phi_1(\br_2) + \phi_0(\br_2)\phi_1(\br_1)\right) \;.
\een
Note that this state is a sum of two Slater determinants. 
\newline
(b) A ground KS singlet state, $\Phi^{\rm gs}$, which, for two electrons corresponds to a single doubly-occupied orbital
\ben
\label{sing}
\Phi^{\rm gs}[n^*](\br_1,\br_2) = {\phi}(\br_1){\phi}(\br_2)\;,
\een
with
\ben
{\phi}(\br) = \sqrt{n^*(\br)/2}
\een
where $n^*(\br) = 2\int \vert \Psi^*(\br,\br')\vert^2 d^3r'$ is the density of the initial excited state.
This is a single Slater-determinant, and is the usual choice when  the true initial state is a ground state.
\newline
(c) A spin-symmetry-broken excited KS state of the form
\ben
\label{spin}
\Phi^*_{\rm SB}(\br'\sigma',\br\sigma) = \frac{1}{\sqrt{2}}\left| \begin{array}{cc}
\phi_0(\br)\delta_{\sigma\uparrow} & \phi_0(\br')\delta_{\sigma'\uparrow} \\
\phi_1(\br)\delta_{\sigma\downarrow} & \phi_1(\br')\delta_{\sigma'\downarrow} 
\end{array} \right|
\een 
This state is not a spin-eigenstate (although it does have $\langle {\bf S}
\rangle = 0$), but is a valid choice for an initial KS state, having
the same total density as the true system. The up and down
spin-densities are not equal, and we shall evolve them in different spin-up and spin-down KS potentials, but,
unlike in spin-DFT, we do not consider the spin-densities separately meaningful;
only their sum is considered as an observable~\cite{DL11}. 
Instead,  (c) will be used to illustrate the effects of orbital-specific functionals, as will be discussed more later.

Now the exact KS potential differs depending on which choice (a), (b), or (c),
is made, but any adiabatic approximation
is identical for them all. Almost all the calculations being run today, certainly all that are coded in the commonly available codes, utilize such an approximation, and insert the instantaneous density into a ground-state approximation:
\ben
v\xc^{\rm adia}[n;\Psi(0),\Phi(0)](\br,t) = v\xc^{\rm gs}[n(t)](\br)\;.
\een
It is not surprising that there are errors inherent in such an
approximation but one question we hope to shed light on by our
investigations here, is how much of this error is due to the lack of
ISD, rather than the lack of history-dependence (dependence on $n(\br,
t'<t)$). Only the latter occurs in usual calculations that begin in an
initial ground-state, where, as mentioned earlier, ISD may be subsumed
into density-dependence.

\subsection{Beginning in the ground-state: ISD or not?}
\label{sec:gsisd}
Many practical situations start with the system in its ground-state;
in fact most calculations, except in the most recent years, have
assumed an initial ground-state. This is fortunate for TDDFT, since,
at least at short times, the adiabatic approximation with its
ground-state functionals, which have become increasingly sophisticated over the years, could then be expected to work reasonably well. 
However,
 we will show that the the subtleties of
initial-state dependence can appear even in this case.

A completely legitimate choice for the KS initial state would be the
true ground-state wavefunction as it trivially satisfies the initial
conditions Eq.~(\ref{eq:n0}) and Eq.~(\ref{eq:ndot0}). The exact KS potential, for this choice, reproduces the density evolution of the exact wavefunction
propagating in the interacting system by propagating the exact
wavefunction in a non-interacting system.

Given that at the initial time we are dealing with a ground-state, one
might expect that the adiabatic approximation would be exact, at least
initially, however this is not the case. The adiabatic approximation
returns the xc potential for a ground-state DFT calculation, where the
KS wavefunction is the ground-state KS wavefunction. If we were to
start the KS calculation with this ground-state KS wavefunction, the
adiabatic approximation is then exact at the initial time.

However the initial
interacting wavefunction is not a ground state of the non-interacting
KS system, and so, choosing this as the initial KS state means the adiabatic approximation will be in error from the
start.

Thus care must be taken when using the short-hand that there is no
initial-state dependence when starting in the ground-state. More
precisely, what is meant is that $1)$ the initial KS orbitals are
chosen to be those of the KS ground-state wavefunction, and $2)$ the
dependence on initial state for this case can be subsumed into the
density via the theorems of ground-state DFT.

\section{Non-interacting electrons}
\label{sec:NIE}

\begin{figure}[t]
    \includegraphics[width=8.5cm,clip]{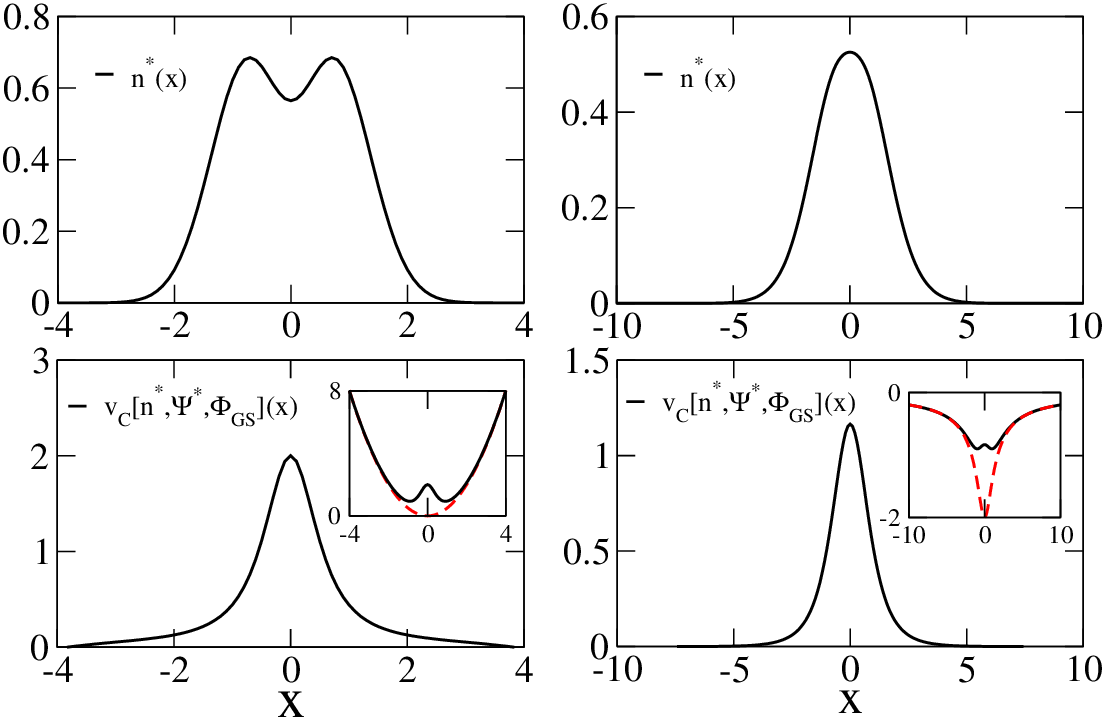}
  \caption{\label{f:NI_vxc}
Top Panels: The first excited-state density, $n^*(x)$, for two non-interacting electrons in a one-dimensional harmonic potential (left) and soft-Coulomb potential (right). Lower Panels: The corresponding correlation potentials, $v\c[\n^*,\Psi^*,\Phi^{\rm gs}[\n^*]](x)$, for an initial KS wavefunction chosen to have ground-state form but yielding the initial excited-state density $n^*$ (case (b) of text). The insert shows the  KS potential (solid line) compared with the external potential (dashed line), chosen as harmonic and soft-Coulomb, respectively. When the initial KS state is chosen to be $\Phi^*$ (case (a) in text),  on the other hand, the correlation potential is zero.}
\end{figure}

The importance of initial-state dependence is strikingly evident even
for the hypothetical case of non-interacting electrons. Due to ISD,
the xc potential is {\it not} always equal
to zero, which may be surprising at first sight, given that there is
no interaction. Although no-one in their right mind would perform
KS calculations for non-interacting electrons for practical
purposes, it is instructive to consider how the KS system behaves in
this case. In particular, the studies suggest what is
the best choice of KS initial state for a given true initial state
when an adiabatic approximation is used.  
Since the adiabatic approximation is designed for ground-states, is the error least if we 
always choose a KS state that is a ground-state?

In the following we consider the ``true'' system to be
non-interacting, i.e. we scale the electron-electron interaction by
$\lambda$ in the limit that $\lambda \to 0$ and consider terms zero-th
order in $\lambda$ only, e.g. the Hartree potential vanishes. We then
consider the xc potential when the KS system is started in different
allowed initial-states.

Consider two non-interacting electrons prepared in an excited state
$\Psi^*$ and evolving in some potential $v\ext(t)$. To start the KS evolution,
any initial state of the same density  $n$, and first time-derivative,
$\dot n$, zero in this case, may be chosen.  We consider the initial state choices (a) and (b) introduced in Sec.~\ref{sec:Back}, which become here: (a) $\Phi(0)
= \Psi^*$, and (b) $\Phi(0) =\phi_0(\br)\phi_0(\br') = \Phi^{\rm gs}, \phi_0=\sqrt{n^*/2}$, where $n^*$ is the density of $\Psi^*$. 

For choice (a), the exact xc potential vanishes, $v\xc[n;\Psi^*,\Phi^*](\br,t) = 0$. 
This can be most easily seen by invoking the uniqueness property of the Runge-Gross mapping: for a fixed particle-interaction, only one potential ($v\ext(t)$) can yield a given density-evolution from a given initial state. Since $\Phi^* = \Psi^*$ for choice (a), we conclude $v\xc(t) = 0$.
This result can  also be seen from the general formula~\cite{L99}:
\ben
\nabla\cdot[n(\br,t)\nabla \left(v\H(\br,t) +v\xc(\br,t)\right)] = q(\br,t) - q\s(\br,t)
\label{eq:divngradvHxc}
\een 
where
\bea
\label{eq:q}
&q(\br,t) &=  \frac{1}{i\hbar}\nabla\cdot \langle \Psi(t) | \ [ \ \hat{\bj}(\br),\hat{T} + \hat{V}\ee \ ] \ | \Psi(t)\rangle \\
&q\s(\br,t) &=  \frac{1}{i\hbar}\nabla\cdot \langle \Phi(t) | \ [ \ \hat{\bj}(\br),\hat{T}\ ] \ | \Phi(t)\rangle
\eea
and $\hat{\bj}(\br)$ is the current-density operator. 
When the two initial wavefunctions are the same,
the RHS vanishes at $t=0$ in the non-interacting limit, and we are left with
$v\H(t=0) = 0$, and $v\xc(t= 0) =0$. Stepping forward in time, it follows that
$v\xc(t) = 0$ for all times.

Now consider choice (b). In this case, $v\xc[n;\Psi^*,\Phi^{\rm
    gs}](\br,0)$ is non-zero: the RHS of Eq.~(\ref{eq:divngradvHxc})
at $t=0$ is no longer zero in the non-interacting limit, due to the
difference in the initial wavefunction in $q$ and $q\s$. That this is
non-zero is also expected by the fact that $v\xc$ must be such that
$v\s = v\ext + v\Hxc$ evolves the initial $\Phi^{\rm gs}$ with the
same density for all time as the different initial state $\Psi^*$ has
when evolved in $v\ext$. So, ISD leads to a non-zero xc potential,
even when the electrons do not interact.

For an explicit demonstration, consider the initial time and realize
that $v\xc(t=0)$ depends entirely on the initial states, as can be
seen from Eq.~(\ref{eq:divngradvHxc}), and not on the choice of the
external potential.  Figure ~\ref{f:NI_vxc} plots this potential
$v\xc(t=0)$ for the first excited state of an external harmonic
potential ($\frac{1}{2}x^2$, on the left), and a soft-Coulomb
potential ($-2/\sqrt{x^2+1}$, on the right) in one-dimension.  The top
panels show the density, and the insets show the external potential in
which $\Psi^*$ is an eigenstate, as well as the the KS potential
$v\ext + v\c$ in which $\Phi^{\rm gs}$ is an eigenstate (the
ground-state, since there are no nodes).  Because for two electrons in
a spin-singlet with one doubly-occupied orbital, $v\x = -v\H/2 = 0$ in
the non-interacting limit, this effect appears entirely in the
correlation potential, and can be interpreted as {\it static
  correlation}: due completely to the non-single-determinantal
structure of the true initial state.  In case (a), this effect
vanishes, because the KS initial state is chosen also not to be a SSD.

We now ask what an adiabatic potential would give: approximating
$v\xc[n;\Psi(0),\Phi(0)]$ by a ground-state potential $v\xc^{\rm
  gs}[n]$.  To distinguish errors arising from the choice of
approximate ground-state functional itself, we consider an
``adiabatically exact'' potential~\cite{HMB02,TGK08},
$v\xc^{\rm adia-ex}[n]$. We shall define this generally in the next
section, but for now it suffices to define it such that if both the
true and KS wavefunctions at all times were in fact ground states of
some potential, then the exact xc potential is the
exact ground-state one, $v\xc[n;\Psi^{\rm gs},\Phi^{\rm gs}] =
v\xc^{\rm adia-ex}[n]$.  Consider again just the initial time, $t=0$. In
the non-interacting limit, for both the true and KS wavefunctions to
be ground-states that have the density of the excited state $\Psi^*$,
then $\Psi(0) =\Phi(0) = \sqrt{n_0(\br)n_0(\br')}/2$, and we rapidly conclude

that 
\ben
v\xc[n,\Psi^{\rm gs},\Phi^{\rm gs}](\br,t) = v\xc^{\rm adia-ex}[n](\br,t) = 0
\een
 in the non-interacting limit. Applying now this adiabatic approximation to the case of the
initial true excited state $\Psi^*$ considered above, we conclude that
{\it the adiabatic approximation is exact for choice (a)}, when the initial
KS state is also chosen excited, as we had argued there the exact
$v\xc$ also vanishes. It is inaccurate for
choice (b), when the initial KS state is chosen to be a ground-state
one, where the exact correlation potential is non-zero.

This study suggests that in the general interacting case, errors in adiabatic
TDDFT will be least when an initial KS state that most closely
resembles the configuration of the true excited state is chosen.  This expectation is indeed borne out in the following studies.

\section{Model Two-Electron Soft-Coulomb Interacting Systems}
\label{sec:SFTC}
The case of non-interacting electrons illustrated the importance of
ISD, while offering a hopeful diagnosis for the adiabatic
approximation; the latter becomes exact if the initial state is chosen
appropriately. Of course, electrons do interact, and now we
turn to studying the effect of ISD on the dynamics of interacting systems. 

When running an approximate TDKS calculation starting in an excited state of the
interacting problem, three separate sources of error come into play. First,
excited states of the exact ground-state KS potential $v\s[n_0](\br)$ do not have the
same densities as interacting excited states of the potential $v\ext(\br)$
whose ground-state density is $n_0(\br)$. 
Yet these are usually the ones chosen in practice. 
We discuss this problem in
Section~\ref{sec:excited}. The second source of error is the central one for
this paper: the use of the adiabatic approximation, when, from the
start, we have an excited state.  We consider first the ``adiabatically exact'' approximation, mentioned also in
Sec.~\ref{sec:NIE}, but considered now for interacting electrons, in
Section~\ref{sec:adia-ex}. This will allow us to separate the errors from the
choice of ground-state functional approximation used in the adiabatic approximation, which is the third issue. 
In Sections ~\ref{sec:E-field},
and~\ref{sec:LiH}, we will 
study the effect that missing initial-state
dependence has in practice, on electron dynamics in model two-electron systems
when begun in the first singlet excited state of the system.

In our model one-dimensional Helium atom, the two
electrons interact via a soft-Coulomb electron-electron interaction,
$v\ee(x_1,x_2) = 1/\sqrt{(x_1-x_2)^2+1}$ and live in an external
soft-Coulomb potential, 
\ben
v\ext(x) = -2/\sqrt{x^2+1} \;.
\label{eq:sftcvext}
\een 
Such a model,
straightforward to solve numerically, is popular in understanding and
analyzing electron interactions, in both the ground-state and in
strong-field dynamics~\cite{sftcreferences}, inside and outside
the density-functional community. In our model of the diatomic LiH molecule, the soft-Coulomb interaction is also used, while the external potential has an asymmetric soft-Coulomb well structure (Sec.~\ref{sec:LiH}).

Before proceding, we make a computational note.  The true
dynamics are propagated on a real-space grid using the exponent
midpoint rule Taylor-expanded to fourth order. Imaginary time
propagation along with Gram-Schmidt orthogonalisation is first
performed to find the lowest eigenstates. The adiabatic exact-exchange
(AEXX) dynamics uses the Crank-Nicolson method with an explicit
predictor step for the Hartree-exchange potential. In both cases, a
timestep of $0.001$ is used with a grid spacing of $0.1$. When
possible these calculations were tested for accuracy against the
parallelized code OCTOPUS\cite{CAOR06}, which was also used for LDA
and LDASIC runs.

\subsection{The Excited-State Density}
\label{sec:excited}

\begin{figure}[t]
    \includegraphics[width=6.8cm,clip]{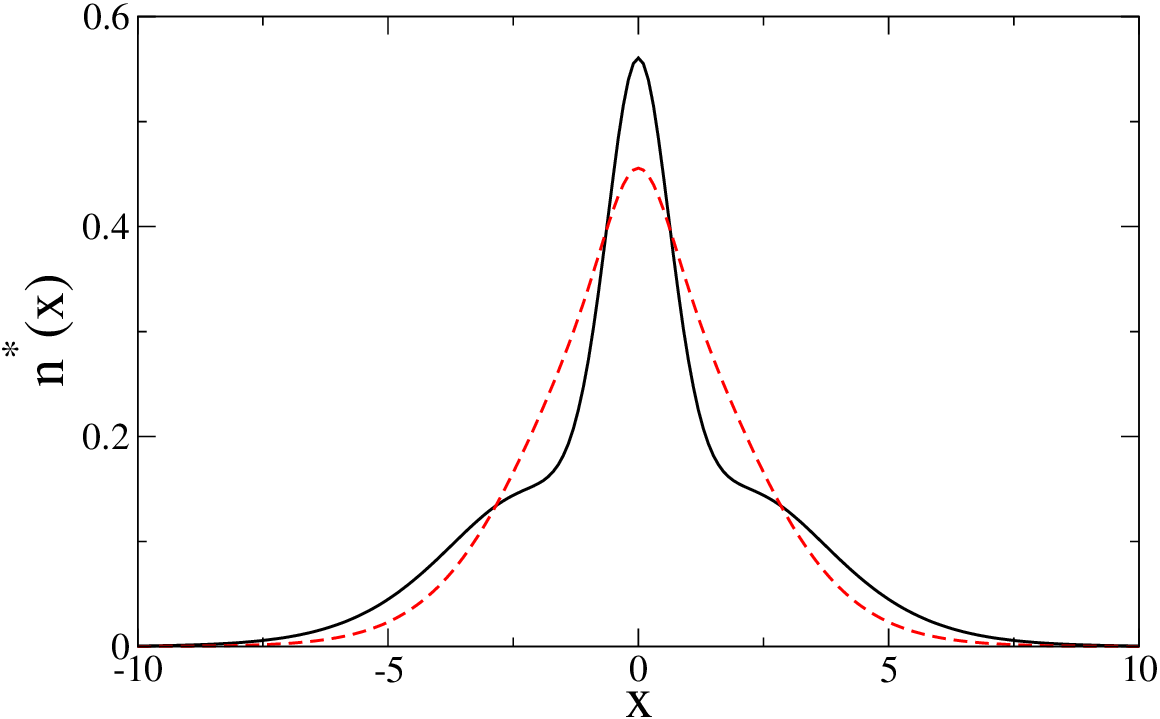}
  \caption{\label{f:SFTC_E1_DEN} Exact excited-state density (solid line) compared to the exact excited-state density of the exact ground-state KS potential (dashed line) for the soft-Coulomb Helium atom.}
\end{figure}

\begin{figure}[t]
    \includegraphics[width=6cm,height=7cm,clip]{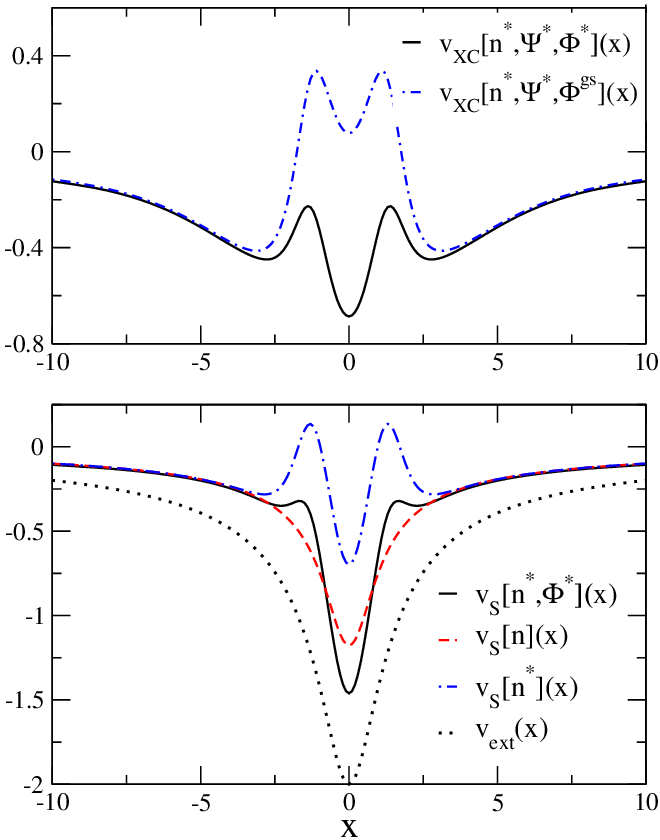}
  \caption{\label{f:SFTC_cmp_vs} Top Panel: The exact xc potentials for the two choices of initial state, $\Phi^*$ (solid line) and $\Phi^{\rm gs}$ (dot-dashed line). Lower Panel: The KS potentials for which the two initial state choices, $\Phi^*$ (solid line) and $\Phi^{\rm gs}$ (dot-dashed line) are eigenstates. Also shown are the exact ground-state KS potential (dashed line) for the soft-Coulomb Helium potential (dotted line).}
\end{figure}

The theorems of TDDFT require that the KS initial state has the same
$\n(\br,0)$ and $\dot{\n}(\br,0)$ as the true interacting system
(Sec.~\ref{sec:Back}). In this section we discuss the difficulty of
fulfilling this requirement when the initial state is not a
ground-state. 

When the true initial state is a ground-state, the natural and usual
choice for the KS initial state is the non-interacting ground-state,
and this can be found by solving the ground-state KS equations.
However, if we want to compute the dynamics of an excited state, we
encounter the problem that there is no DFT scheme to find excited
state densities\cite{GB04}. Furthermore, even if we use a more
computationally expensive higher-level wavefunction method to
calculate the state $\Psi(0)$, we still must then choose an initial
$\Phi(0)$ in which to start the KS calculation. If we start the
interacting system in a first excited state of some $v\ext(\br)$, the
results of the earlier sections suggest a good choice for adiabatic
TDDFT is to start the KS system in a corresponding non-interacting
excited state where one electron is excited from the highest occupied
to lowest unoccupied orbital of some $\tilde v\s$ (as in form (a) of
Sec.~\ref{sec:Back}). This $\tilde v\s(\br)$ however cannot be the
ground-state KS potential $v\s(\br)$ corresponding to the interacting
$v\ext$, because this would not satisfy Eq.~(\ref{eq:n0}): $v\s(\br)$
yields the same ground-state density in a non-interacting system that
$v\ext(\br)$ yields in the interacting one, but the densities of their
excited-states are different.  Yet, the usual practice {\it is} to use
the corresponding excited-state of the ground-state KS potential~\cite{TRR05}. This
is partly because the density of the interacting excited state is not
often known anyway.
In Fig. \ref{f:SFTC_E1_DEN}
we compare the exact density of the first excited interacting singlet state in the soft-Coloumb He atom (Eq.~\ref{eq:sftcvext}) with the density of the excited state of the corresponding ground-state  KS potential. 
Although matching the
general shape, the latter is too narrow and misses some structure.

To be able to separate the error from not quite having the exact
initial density from the error from using an adiabatic approximation,
we now search for a non-interacting system where the first singlet
excited-state density {\it is} exactly equal to the true excited-state
density $n^*$, for our model He atom.  Several such potentials may
exist~\cite{GB04}, and
in the lower panel of Fig.~\ref{f:SFTC_cmp_vs}, we plot one such KS potential, denoted
$v\s[\n^*,\Phi^*](x)$ and compare to the  KS potential
where $\n(x)$ is the exact ground-state density $v\s[\n](x,0)$ of the model He atom. (The
first excited state densities of these potentials are precisely those
shown in Fig. \ref{f:SFTC_E1_DEN}). 
We 
also show $v\s[\n^*](x,0)$, i.e. the KS potential for which $\n^*(x)$ is the density of the ground-state. This corresponds to choice (b) in Sec.~\ref{sec:Back} for the initial KS state.

Once a valid KS state is found satisfying Eqs.~(\ref{eq:n0})
and~(\ref{eq:ndot0}), the initial xc potential is completely
determined: Eqs.~(\ref{eq:divngradvHxc}) and ~(\ref{eq:q}) show that at the
initial time, $v\xc$ is a functional of just the two initial states
$\Psi(0)$ and $\Phi(0)$. This can be added to any external potential,
prescribed by the physical problem at hand, and Hartree-potential,
determined by the initial density, to start the evolution. In other words,
the choice of the initial KS state fundamentally points to an xc
potential, rather than to a KS potential, and so in the top panel of Fig.~\ref{f:SFTC_cmp_vs} we plot the xc potentials corresponding to choices (a) and (b) of the initial KS state.
In the following examples, we take the initial $v\ext$ to be
that of the soft-Coulomb Eq.~(\ref{eq:sftcvext}); sometimes subsequently an 
external field is added.

We have therefore, now found initial states of the form (a) and (b) of
Sec~\ref{sec:Back}, both yielding the same density as that of the true
excited-state, meeting the conditions Eq.~(\ref{eq:n0})
and~(\ref{eq:ndot0}).  We can now move forward to investigate the
time-dependent properties.

\subsection{The Exact Adiabatic Approximation}
\label{sec:adia-ex}

\begin{figure}[t]
    \includegraphics[width=6cm,height=7cm,clip]{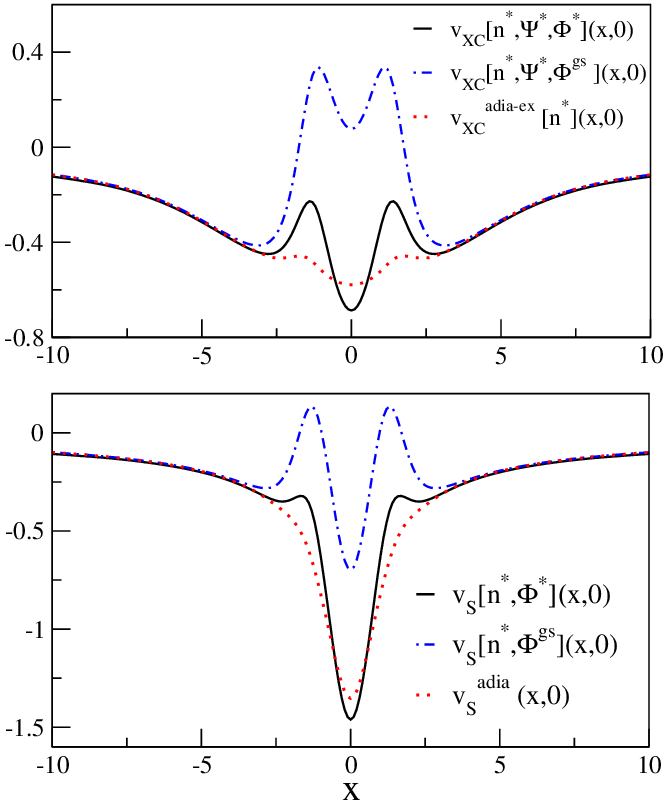}
  \caption{\label{f:SFTC_cmp_vs_adia} Top Panel:The exact xc potentials and the exact adiabatic xc potential. Lower Panel: The exact and adiabatically-exact KS potentials.}
\end{figure}

\begin{figure}[t]
    \includegraphics[width=9cm,clip]{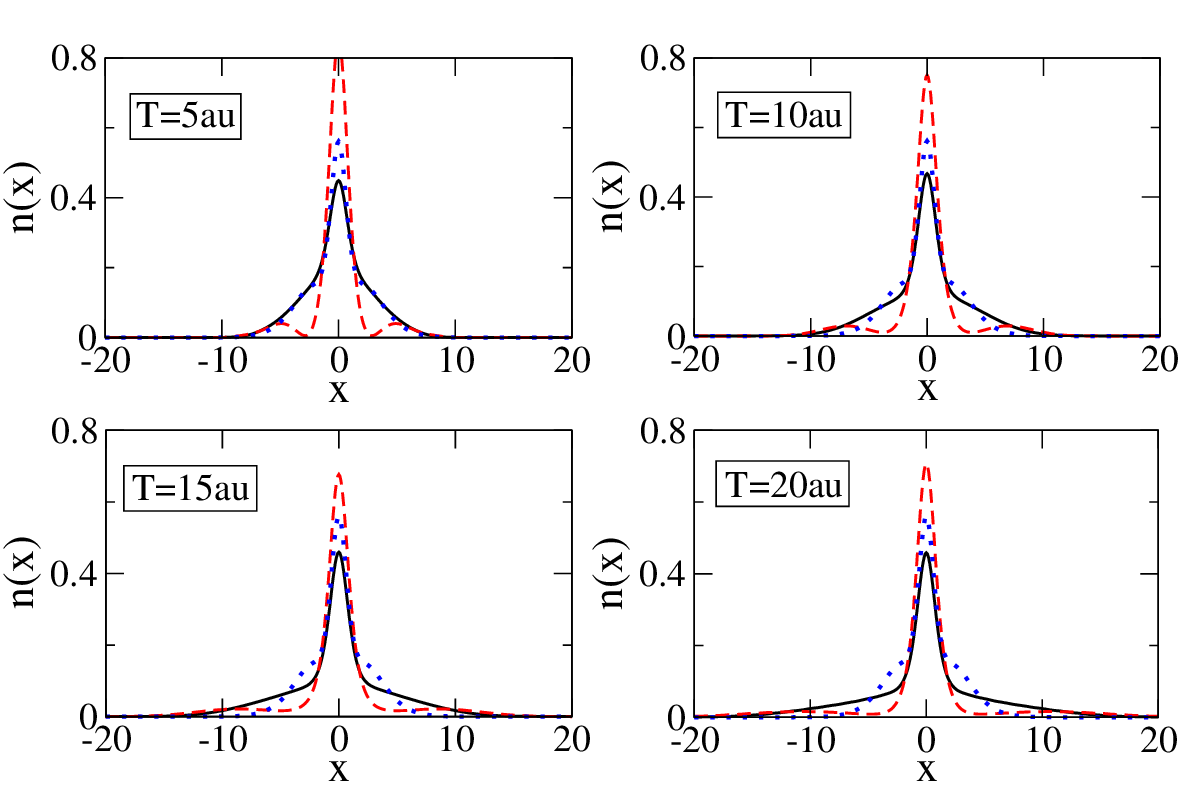}
  \caption{\label{f:SFTC_EXPHI_ADIA_DEN} The density at times $5$au, $10$au, $15$au, and $20$au for initial states $\Phi^*$ (solid line) and $\Phi^{\rm gs}$ (dashed line) propagating in the static exact-adiabatic initial potential. Also shown is the exact excited-state density (dotted line) which is the exact solution at each time.}
\end{figure}

We next move to the error introduced by using the adiabatic approximation.

To isolate this error we will calculate the adiabatically exact
potential at the initial time, and start in a KS state of the exact same density as the
true state, as found in Section~\ref{sec:excited}.

The adiabatically exact potential is defined by~\cite{TGK08} 
\ben
v\xc^{\rm adia-ex}[\n] =
v\s^{\rm adia-all}[\n] - v\ext^{\rm adia-all}[\n] -v\H[\n]
\label{eq:adiaex}
\een
where $v\s^{\rm adia-all}[\n](\br,t)$ is the potential for which $\n(\br,t)$ is
the non-interacting ground state density and $v\ext^{\rm adia-all}[\n](x)$ is
the potential for which interacting electrons have $\n(\br,t)$ as their
ground-state density. If both the true and KS wavefunctions are actually always in some ground-state, then Eq.~(\ref{eq:adiaex}) becomes the exact xc potential.
We shall consider this potential for initially excited states, only at the initial time.

The non-interacting potential
$v\s^{\rm adia-all}[\n^*](\br,0)$ may be easily found by inverting the KS
equation for a doubly-occupied orbital, ${\phi}(\br) = \sqrt{\n^*(\br)/2}$ where $n^*(\br)$ is the density of the initial state $\Psi^*$.

The interacting potential, $v\ext^{\rm adia-all}[\n](\br,0)$ is
solved for using an iterative technique whereby the external potential is
updated based on the difference between the ground-state density of
the current iteration and the density we are targeting. This is based
on the inversion algorithm of Ref. \onlinecite{PNW03},  but generalized to the interacting case\cite{TGK08}. We also increase the update to the potential in regions of low density by simply using the inverse of the density (up to a maximum value) as a weighting factor, similar to Ref. \onlinecite{DL11}. When the density converges to the target
density, we have found $v\ext^{\rm adia-all}[\n](x)$.

The dotted line in the lower panel of Fig. \ref{f:SFTC_cmp_vs_adia},
shows at the initial time, the full KS potential with the
adiabatically exact xc potential, namely $v\s^{\rm adia}(x) = v\ext(x)
+ v\H[\n](x) + v\xc^{\rm adia-ex}[\n](x)$, choosing $v\ext(t=0)$ as
the soft-Coulomb potential of Eq.~\ref{eq:sftcvext}, for the first
excited singlet state $\Psi^*$. Alongside, we compare this to the
exact KS  potentials found in Sec.~\ref{sec:excited} (see
Fig. \ref{f:SFTC_cmp_vs}), for the initial state choices of case (a)
and (b) respectively.  The adiabatically exact KS potential $v\s^{\rm
  adia}(x)$ tracks the shape of the exact KS potential
$v\s[n^*,\Psi^*,\Phi^*]$ although misses structure. The difference
between them is the difference in their xc potentials shown in the upper panel of
Fig. \ref{f:SFTC_cmp_vs_adia}.
The difference is much larger for the comparison with the exact KS
potential when the initial KS state is chosen as the ground-state,
(i.e.  $v\xc^{\rm adia-ex}[n^*]$ is closer to $v\xc[n^*,\Psi^*,\Phi^*]$  than it is to $v\xc[n^*,\Psi^*,\Phi^{\rm gs}[n^*]]$); in this case, there is a large
static-correlation contribution to the exact xc
potential. The results here are consistent with the non-interacting 
case discussed in  Sec.~\ref{sec:NIE}: with interaction, the adiabatic
approximation is no longer exact for the choice of $\Phi^*$, however it is still
the better choice over $\Phi^{\rm gs}$.

The fact that the adiabatically exact potential does not equal the
true potential, even at the {\it initial} time, leads to erroneous
dynamics.  To illustrate this effect we  propagate our exact
initial wavefunctions in the adiabatically exact potential and compare
with exact dynamics.  Since calculating at each time is
computationally demanding and delicate, we will simply  evolve in time without any additional perturbing
potentials, and hold the adiabatic potential fixed to its initial
value. The idea is that since the exact evolution is static, the
adiabatically exact potential is static and equals its initial value
at all times; if this was a good approximation, it would yield
density-dynamics that were close to static. The deviation of the
adiabatic propagation from the exact is a measure of its error.
Note such a calculation does not treat the adiabatic potential
self-consistently; later calculations with approximate adiabatic functionals suggest having self-consistency decreases the error somewhat.

In Fig. \ref{f:SFTC_EXPHI_ADIA_DEN}, we plot snapshots of the density
at $5$au intervals until $T=20$au for the two choices of exact initial
states discussed above, evolving in the fixed
adiabatically exact potential plotted in Fig.~\ref{f:SFTC_cmp_vs_adia}. As
each time we show the exact static excited-state density. While both choices
show density 'leaking' out, the $\Phi^*$ choice is more successful in
preventing this, as might be expected from the above discussions, but cannot stop the density melting away in its outer
regions. The $\Phi^{\rm gs}[n^*]$ choice is poor from the start, jettisoning
density outwards and becoming too narrow.

This error in the adiabatically exact evolution is caused {\it entirely} by
initial-state dependence: for the analogous calculation
beginning in the interacting ground-state and starting with the ground-state KS
wavefunction, the adiabatic approximation would be exact at all times
when there are no perturbing fields.

\subsection{Approximate GS functionals}

\begin{figure}[t]
    \includegraphics[width=6.8cm,clip]{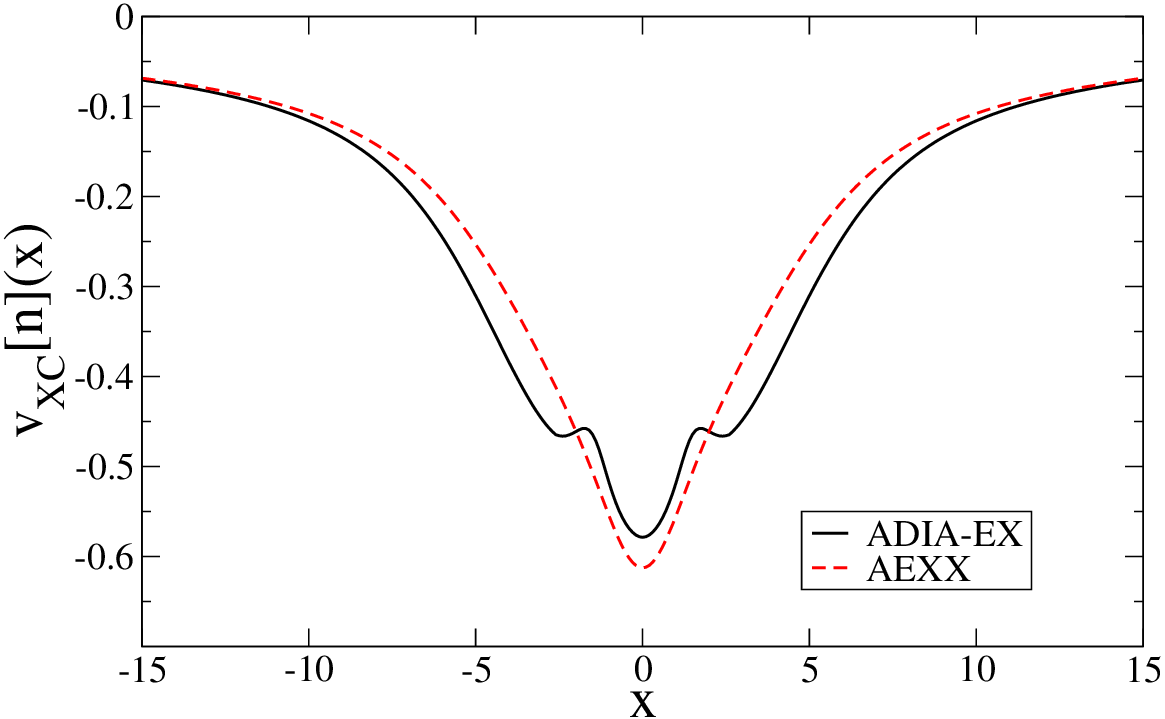}
  \caption{\label{f:SFTC_vxc_Ads} The exact adiabatic xc (solid) and AEXX (dashed) potentials evaluated on the true excited-state density $n^*$. (Note that the solid line here is the same as the dotted line in the top panel of Fig. 4}
\end{figure}

\begin{figure}[t]
    \includegraphics[width=7cm,clip]{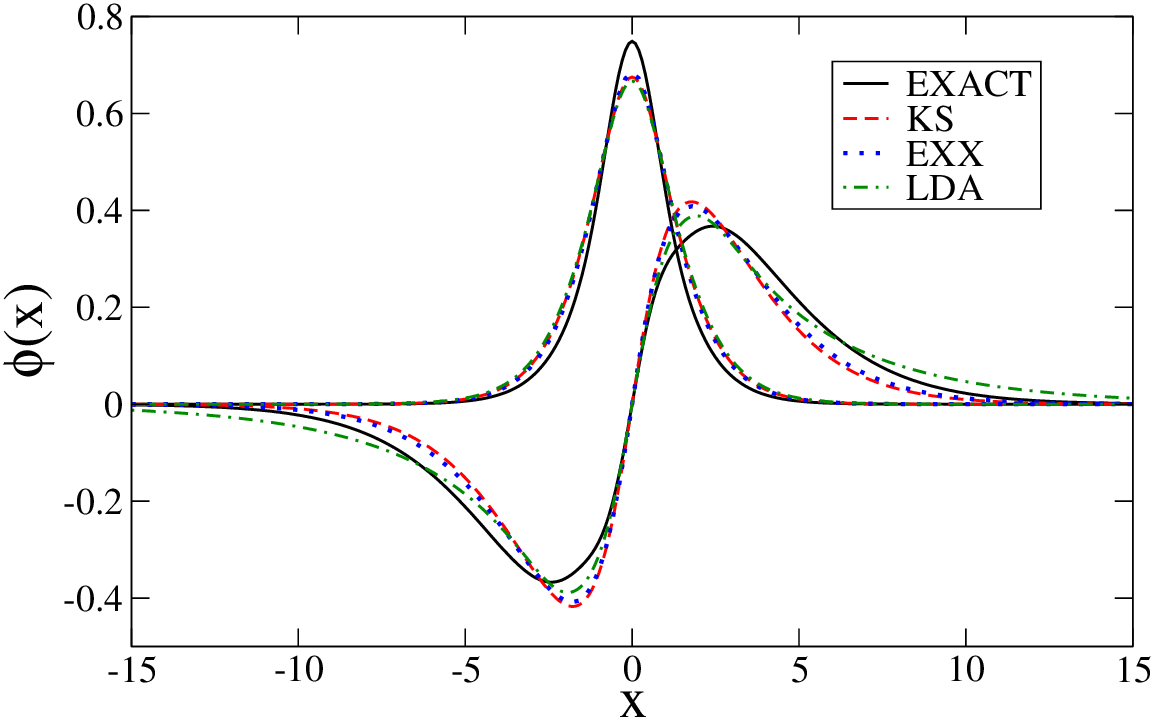}
  \caption{\label{f:SFTC_cmp_phi} The ground and first excited orbitals from ground-state EXX and LDA, compared to those from the exact GS KS potential. Also shown are those orbitals found in Sec. \ref{sec:excited} of this configuration that yield the exact excited-state density.}
\end{figure}

Finally we look at the third source of error, namely using an
approximate ground-state xc functional instead of the exact one.  This
error contributes at both the ground-state level in obtaining the
initial orbitals, and in the adiabatic functional when computing
dynamics. We first turn our attention to the latter and investigate
the adiabatic exact-exchange functional, AEXX, which, for a
two-electron system, is simply $v\xc^{\rm AEXX}[\n](\br,t) =
-v\H[\n](\br,t)/2$. In Fig. \ref{f:SFTC_vxc_Ads} we compare the
adiabatically exact xc potential found in the previous section to AEXX,  at the initial
time.  The undulations of the exact adiabatic xc potential that are
missing in the AEXX represent correlation; they are relatively small
in this particular case once added to the external potential, so we expect
that AEXX dynamics fairly approximates the adiabatically exact dynamics in this situation at least for short times. 

We now return to the error of not having the correct initial density (Sec.~\ref{sec:excited}). 

In a usual TDDFT calculation, a ground-state DFT calculation is first
performed and the orbitals from this are used to create the initial
wavefunction. So we would start with the ground-state KS density of
Fig. \ref{f:SFTC_E1_DEN} except here there is a further error that will be
made, as an approximate ground-state xc
functional is used instead of the exact one.

In Fig. \ref{f:SFTC_cmp_phi} we plot the ground- and first-excited- KS
orbitals from ground-state DFT calculations using EXX and LDA for the
soft-Coulomb helium atom, and compare to the exact KS orbitals. We
also plot, for completeness, the pair of orbitals found for this
configuration that yield the exact {\it interacting} excited-state density,
i.e. these are the exact orbitals which, when singly-occupied, make up
the densities shown in Fig. \ref{f:SFTC_E1_DEN}.

If we first compare the approximate orbitals to the exact KS orbitals
of the soft-Coulomb potential, we find both LDA and EXX perform very
well, particularly for the ground-state orbital. Their ground-state KS
potentials are however quite different: it is well known that the LDA
potential is much too shallow, with a much too rapid exponential decay
away from the nucleus, while the EXX is much closer to the exact, with
the correct asymptotic decay. This is reflected in the first-excited
orbital in LDA, which is more diffuse than EXX and the first-excited
orbital of the exact soft-Coulomb KS potential.  The asymptotic
behavior will be important especially when we turn on
an electric field, so we will (mostly) use the EXX orbitals to build our
initial KS wavefunctions.

In the next section we will add the spin-symmetry-broken initial state
(c) introduced in Section~\ref{sec:Back} to our investigations; while
no longer a spin eigenstate like the exact case or $\Phi^*$ and
$\Phi^{\rm gs}$, this still has the same total density as the exact
system. We will simply construct this from the same EXX orbitals
composing $\Phi^*$, but will evolve the two orbitals using {\it
  different} xc potentials, $v\xc^\uparrow$ and $v\xc^\downarrow$.  For exact-exchange, we have
$v\Hxc^{\uparrow\downarrow}[\n](x) = v\H[\n^{\downarrow\uparrow}](x)$
for two electrons.  We are however {\it not} doing spin-DFT since the
individual spin-densities are not physical ones; we only ever consider
their sum as observable. (Indeed, even the initial spin-densities are
wrong). The point of considering such an initial state and dynamics is
that, for two electrons, it is an example of dynamics in {\it
  orbital-specific} potentials, as would occur in generalized KS
approaches~\cite{SGVM96,K12,GL97}. Orbital-dependent
functionals require an optimized effective potential approach to find
a single potential for all the KS orbitals to live in. This procedure
is numerically very intensive, so often a generalized KS approach is
used, that relaxes the condition that all orbitals evolve under the
same potential; moreover, the latter has shown to have certain
advantages over the OEP in certain cases, e.g. better band-gaps.  The
interest in orbital-dependent functionals is that that they can work
less hard to capture both spatially-non-local and time non-local-
density-dependence since the orbitals themselves are non-local
functionals of the density.

We note nevertheless that it can be shown there is no ISD for
two-electron dynamics with spin-TDDFT; this follows straightforwardly
generalizing the arguments of Ref. \onlinecite{MB01} to each
spin-density.  This is not the case for our spin-broken case, however,
as here we are {\it not} attempting to reproduce the exact
spin-densities but rather the total density evolution; only $n(\br) =
n_{\uparrow} + n_{\downarrow}$ will be considered meaningful.

Having delineated and explored the three possible sources of error in
an adiabatic TDDFT calculation of excited state dynamics, we are now
ready to run a typical adiabatic TDDFT calculation on our model
systems, starting with our different choices of initial KS states, and compare with the exact result.

\subsection{Propagation in an electric field}
\label{sec:E-field}
\begin{figure}[t]
    \includegraphics[width=7cm,clip]{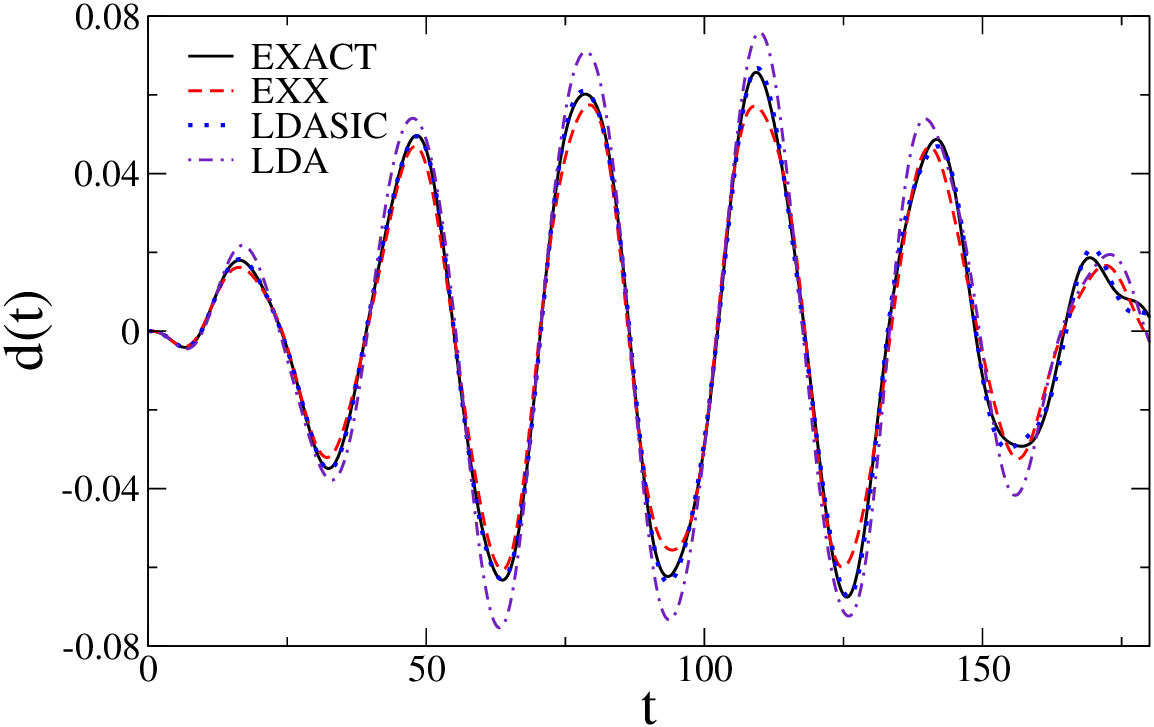}
  \caption{\label{f:SFTC_gs_dip} The exact dipole moment and those from TDDFT calculations starting in the self-consistent ground-state KS wavefunction and propagated with the corresponding adiabatic approximation, all under the influence of the electric field described in the text.}
\end{figure}

\begin{figure}[t]
    \includegraphics[width=7cm,clip]{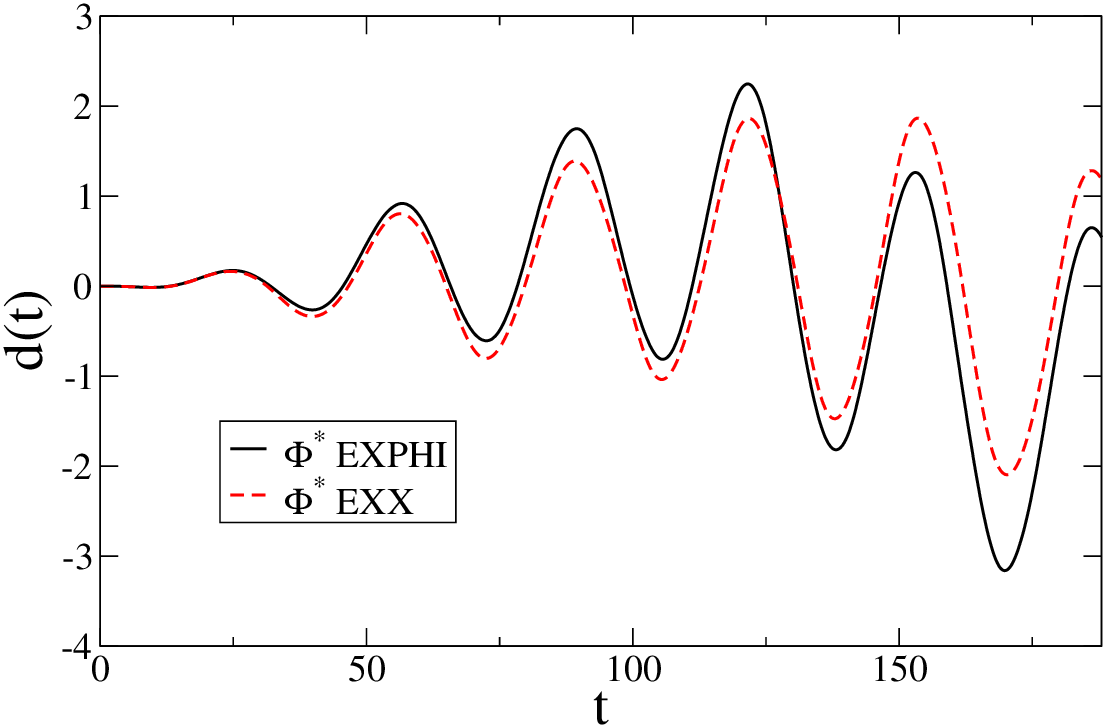}
  \caption{\label{f:SFTC_ETRAP_PHIS_d} The dipole moments when beginning in $\Phi^*$ composed of EXX orbitals (dashed line) and the exact orbitals of Sec. \ref{sec:excited} (solid line) in the electric field (see text).}
\end{figure}

\begin{figure}[t]
    \includegraphics[width=7cm,clip]{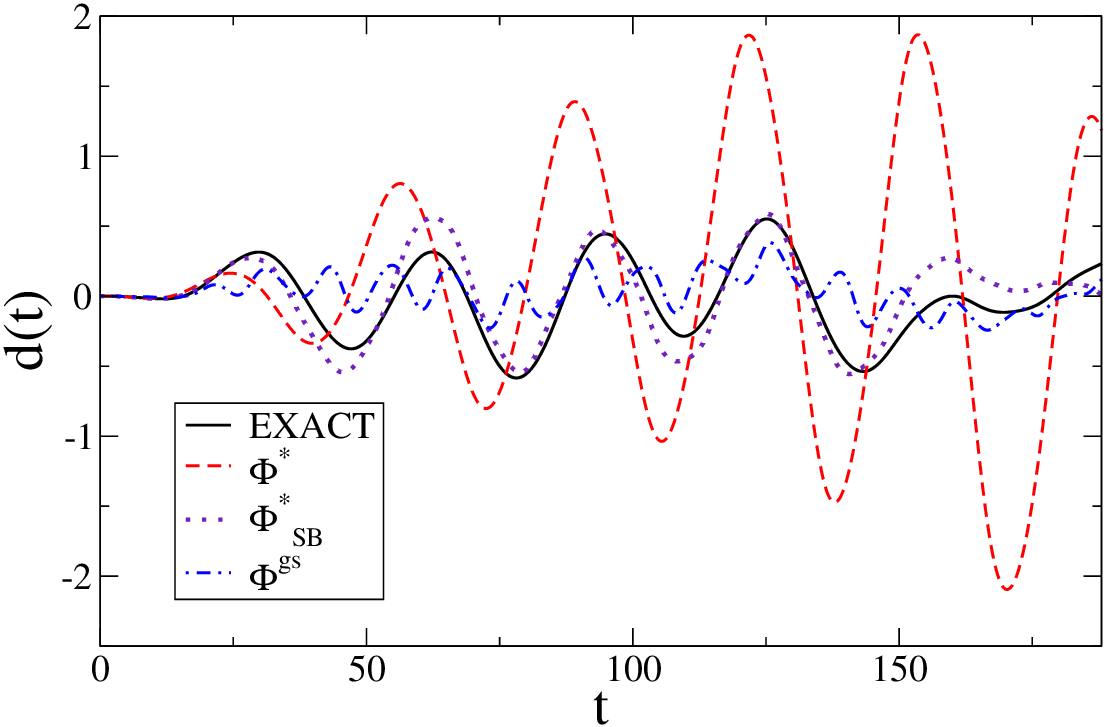}
  \caption{\label{f:SFTC_ISdip} The dipole moments for dynamics in the given electric field starting in the $\Phi^*,\Phi^*_{\rm SB},\Phi^{\rm gs}$ initial KS wavefunctions and using AEXX, compared to exact.}
\end{figure}

\begin{figure}[t]
    \includegraphics[width=7cm,clip]{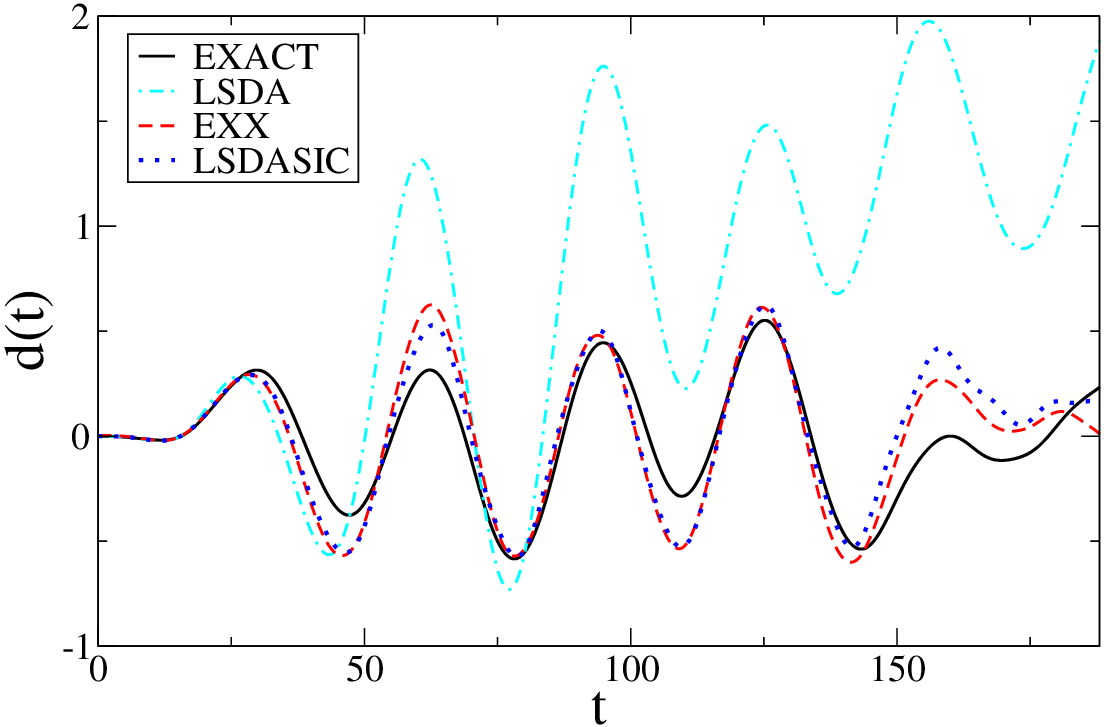}
  \caption{\label{f:SFTC_E1_SPIN} Dynamics for the spin-broken case, $\Phi^*_{\rm SB}$, for different orbital-specific adiabatic functionals, EXX, LSDASIC, and LSDA, compared to the exact.}
\end{figure}

We simulate electric-field driven dynamics in the soft-Coulomb helium
model, using adiabatic TDDFT. We apply a relatively weak oscillating electric field field of
amplitude $0.01$au with an off-resonant frequency of $0.2$a.u. We run
for $6$-cycles including a trapezoidal envelope consisting of a
$2$-cycle linear switch on and a $2$-cycle switch off.

As a point of reference, we first look at the performance of TDDFT
when starting in the {\it ground state}. We propagate the
doubly-occupied EXX ground-state orbital in the electric field, using
AEXX, and compare to the exact dynamics of the initial interacting
ground-state. The resulting dipole moment is very accurate as can be
seen in Fig. \ref{f:SFTC_gs_dip}. In fact, almost exact dynamics can
be achieved in this case with an adiabatic approximation: the same
figure shows the result of using the adiabatic self-interaction
corrected LDA approximation (LDASIC) to propagate the LDASIC
ground-state orbital. The dipole moment lies practically on top of the
exact curve. Including correlation in a self-interaction free functional therefore improves over exact
exchange, but an adiabatic approximation is certainly adequate in this case. We also show for comparison  the LDA result.

The adiabatic approximation is however not so rosy, as we shall see, in the case of an
initially excited state.  The initial interacting state is the first
singlet excited state ($E=-1.705$au). 

The typical TDDFT calculation will begin in an excited state
determined by the orbitals of the corresponding ground-state KS
system, but as discussed in Sec.~\ref{sec:excited}, this does not have
quite the right density to start with.  Since in our model system, we
found an initial KS state of the correct density, we first check the
error that using the AEXX orbitals make.  We plot in
Fig. \ref{f:SFTC_ETRAP_PHIS_d} the dipole moments for AEXX
calculations starting in the $\Phi^*$ initial state with exact
orbitals and with the ground-state EXX orbitals (i.e. using the solid
and dotted orbitals of Figure~\ref{f:SFTC_cmp_phi} respectively).  As

previously anticipated, the two calculations
perform  similarly, especially at shorter times.

At last we are ready to study the excited-state electron dynamics,
using, as in practical calculations, an approximate ground-state KS
potential (in this case EXX) to generate the orbitals composing the
various initial state choices.  In Fig. \ref{f:SFTC_ISdip}, we plot
the dipole moment for each choice of initial state. The
spin-broken-symmetry case gives the best dynamics; as mentioned
earlier, functionals expressed directly in terms of instantaneous
orbitals automatically contain time non-local and spatially-non-local
density-information. In this particular case the orbital-specific EXX
potentials, obtained from spin-scaling the (spin-restricted) AEXX
potential, are self-interaction-free, while the latter is not for the
case of the {\it excited} two-orbital singlet state.

 Unlike in the propagation of the ground-state, adding correlation to
 the adiabatic potential does not significantly improve the results,
 as can be seen in Fig. \ref{f:SFTC_E1_SPIN}; the symmetry-broken
 LSDASIC and EXX calculations are both off by roughly the same amount. Fig. \ref{f:SFTC_E1_SPIN} also shows the LSDA result, clearly worse than the others by comparison.

The $\Phi^*$ initial state has the correct spin symmetry and is
closest in character to the true excited state, being a double Slater
determinant. It gives better dynamics than the $\Phi^{\rm gs}[\n^*]$
initial state, whose dynamics are completely incorrect. This is
consistent with the results in the previous sections. We can say with
confidence that it is ISD, and its underlying static correlation that
it causing poor dynamics for this doubly-occupied-singlet case.

We can understand the poor result for the initial state of the
Eq.~(\ref{sing}) form also by looking at the exact adiabatic xc potential
and $v\xc[\n^*,\Psi^*,\Phi^{\rm gs}[n^*]](x)$, both shown in
Fig. \ref{f:SFTC_cmp_vs_adia}. This form treats the density as if it
were a ground-state and uses one doubly occupied orbital, inversion of
this orbital yields $v\s^{\rm adia-all}[\n](x)$, i.e. this is the
potential in which the wavefunction is static at the initial
time. However in a calculation using the adiabatic approximation , it
will fall into a much deeper potential, drastically effecting its
dynamics. Without the initial-state dependence of the xc potential
giving a bump similar to Fig. \ref{f:NI_vxc}, it will perform poorly.

\section{Model LiH Dynamics}
\label{sec:LiH}
Our final example is perhaps the most topical one where the initial
state is an excited one: coupled electron-ion dynamics after
photo-excitation. The initial excitation is assumed to place the
initial nuclear wavepacket on an excited potential energy surface,
vertically up from its ground-state equilibrium. Field-free dynamics
on the excited surface ensues; usually the nuclei are treated
classically, coupled in either an Ehrenfest or surface-hopping scheme
to the quantum electron dynamics.  For Ehrenfest dynamics in the TDDFT
framework~\cite{TRR05}, and the simplest type of
surface-hopping~\cite{CDP05}, the electrons start in the KS
excited-state obtained simply from promoting an electron from the
highest occupied orbital in the ground-state KS configuration to a
virtual orbital.  This initial state is then evolved in the
time-dependent external potential caused by the moving ions. The
dynamics of the ions is given by simple Newtonian mechanics where the
electronic system provides a force $\int n(\br,t)\nabla_R v\ext(\br;\{ {\bf R} \})$, where $\{ {\bf R} \}$ represent the nuclear coordinates.  All
the three errors mentioned earlier therefore raise their heads: the
initial density is not that of the true excited state (even if the
exact xc potential was used), an adiabatic approximation is used to
propagate it, and this involves an approximate ground-state
functional.  It should be noted that in the more accurate
surface-hopping method of Ref.~\cite{TTR07}, where
correct TDDFT surfaces are used to provide the nuclear
forces~\cite{M06}, the first problem does not arise; the method in
fact circumvents having to define the electronic state.

We will use the parameters found in Ref. \onlinecite{TMM09} for a one-dimensional two-electron lithium hydride model, where 
\ben
v\ext(x) = -1/\sqrt{x^2+0.7} -1/\sqrt{x^2+2.25}
\een
 and $v_{nn}(R) = -1.0/\sqrt{R^2+1.95}$.
We integrate Newton's equations of motion
using the leapfrog algorithm, and for simplicity, we use
the same time step for both electrons and ions.

\begin{figure}[t]
    \includegraphics[width=7cm,clip]{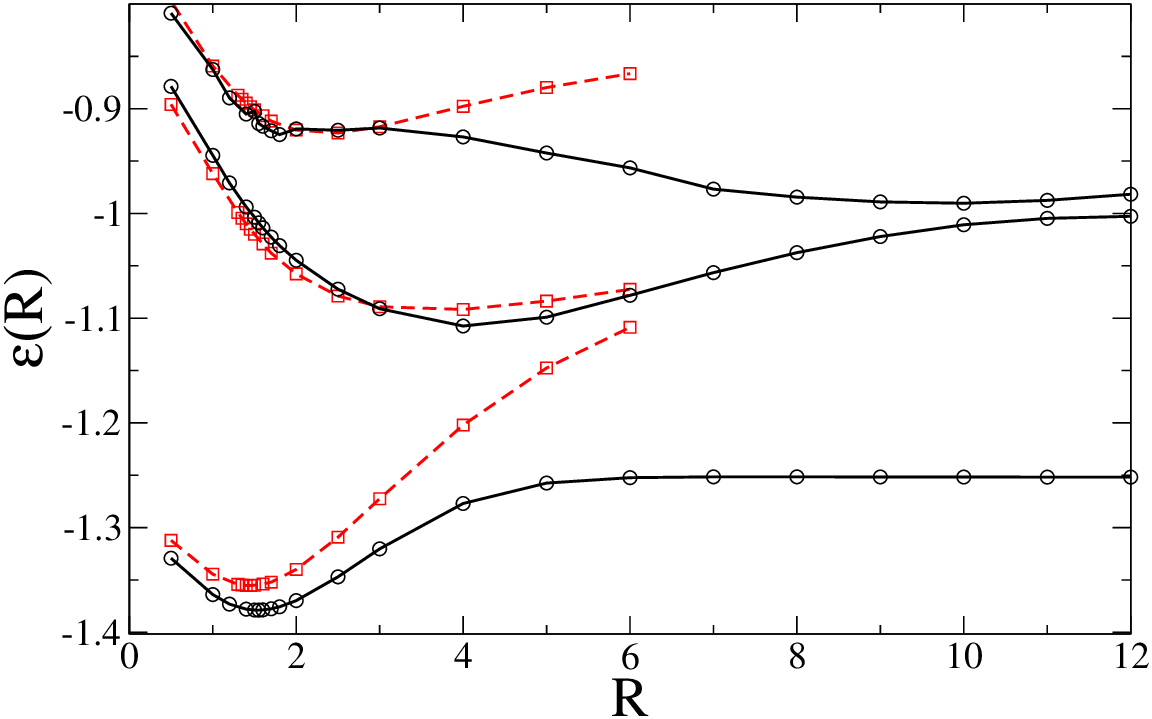}
  \caption{\label{f:LiH_BO} The exact Born-Oppenheimer potential energy surfaces (solid lines) and the EXX KS Born-Oppenheimer surfaces (dashed lines). Note that due to degeneracy between the lowest two surfaces, these calculations cannot be converged for large values of R, the interatomic separation.}
\end{figure}

The Born-Oppenheimer potential energy surfaces are shown in
Fig. \ref{f:LiH_BO} for both the exact system and those of the bare KS system calculated within 
spin restricted EXX. 

The ground state surface has a minimum at
$R=1.55$, whereas the EXX has the gs equilibrium at $R=1.45$.
Near equilibrium the surfaces are somewhat similar.
 The spin-restricted EXX surfaces however encounter the
well-known fractional charge problem as $R$
increases, with the calculations becoming extremely difficult to
converge and the lowest two surfaces collapsing onto each
other.
 In the
exact surface, an avoided crossing can be seen between the ground and
first excited state around $R=4$, where the latter establishes ionic character, eventually
decaying as $-0.896-1/x$, while the neutral dissociation curve flattens out. In
fact there are multiple avoided crossings at larger distances, the
next being around $R=10$. Unrestricted EXX calculations break the spin-symmetry of the true
ground-state at a critical separation, but describe dissociation and the shape of the potential energy surfaces at larger distances qualitatively~\cite{CGGG00,FRM11}.

\begin{figure}[t]
    \includegraphics[width=7cm,clip]{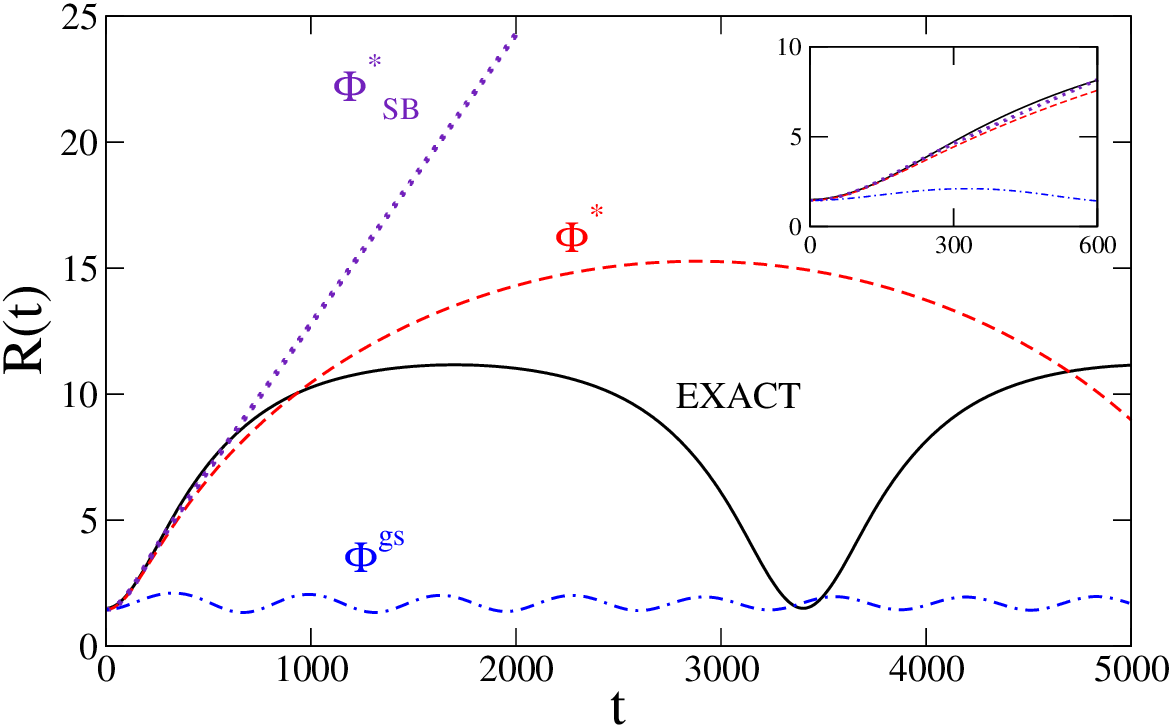}
  \caption{\label{f:LiH_coord} The interatomic distance, $R(t)$, for Ehrenfest dynamics starting in the first excited-state and propagated freely. The various KS initial wavefunctions ($\Phi^*,\Phi^*_{\rm SB},\Phi^{\rm gs}$) are propagated with the adiabatic exact exchange approximation. Insert: short-time dynamics for the various initial states.}
\end{figure}

We vertically excite the system at the ground-state equilibrium
geometry into the first excited singlet state and then begin the
propagation with the initial nuclear momentum chosen as zero. In
Fig.~\ref{f:LiH_coord}, we show the internuclear separation as a
function of time for our three choices of initial states, calculated
within EXX. For $\Phi^*$ we make the usual practise of taking $\phi_0$
in Eqs.~\ref{dsd} and \ref{spin} as the HOMO and $\phi_1$ as the LUMO; likewise
for the spin-symmetry-broken version $\Phi^*_{\rm SB}$. The density of
this state $n^*$ is then used to define the doubly-occupied orbital in
the SSD state $\Phi^{\rm gs}[n^*]$ (Eq.~\ref{sing}).

For this particular system, the exact system remains on the first
excited surface; the nuclear coordinate oscillates between the turning
points at $R=1.5$ and $R=11.165$.  The ground state $\Phi^{\rm gs}$
completely fails to capture this behavior, making only very small
oscillations around its initial position. The excited states $\Phi^*$
and $\Phi^*_{\rm SB}$ show reasonable agreement for short times, with
the spin-broken state once again edging out the other. For longer
times however, the spin-broken state is dramatically wrong, yielding
nuclei moving away from each other with constant speed. The excited
state $\Phi^*$ shows behavior more qualitatively exact but, because of
the deviation of its potential energy curve compared to the exact one,
the oscillation period is quite wrong.

The conclusions regarding ISD in this example, at least at short
times, are consistent with the results throughout this paper and
support the idea that when using an adiabatic approximation choosing a
KS initial-state with a configuration close to that of the exact
interacting initial state yields the better result. However, the
behavior at longer times in this problem highlights, above all, the
need for accurate ground-state functionals: EXX produces poor nuclear
dynamics when $\Phi^*$ is chosen as the KS state primarily because the
shape of its potential energy surfaces is wrong -- the ground state as
well as the excited state ones. Any adiabatic approximation that does
not have strong correlation will not perform well. Despite giving
improved potential energy surfaces, the spin-broken EXX approach fails
for dynamics for longer times. We believe this is because as the
molecule dissociates, symmetry-breaking localizes each electron on one
nucleus or the other, with each evolving according to a different xc
potential; each atom sees just a neutral atom, leading to a net zero
potential, and a constant velocity (somewhat as if the molecule was
dissociating on the ground-state surface).  It should be noted that
the spin-broken solution is not evolving on the spin-broken
PES. e.g. even at the initial time, at the equilibrium geometry there
is no symmetry-breaking in a spin-unrestricted approach, so the
spin-broken surface is on top of the spin-restricted surface showed in
the figure. However the spin-broken wavefunction (c) is evolved using
different potentials for each orbital, unlike the evolution dictated by either the spin-restricted EXX surface or the unrestricted EXX (the symmetry-breaking point occurs around R = $2.5$). 

\section{Conclusions and Outlook}
\label{sec:Con}
Until relatively
recently, ISD fell in the realm of a theoretical curiosity, but due
to an increasing number of topical applications beginning with the system
not in its ground-state, ISD is now also of much practical concern.
Almost all calculations today use adiabatic functionals that neglect
the ISD that the exact functionals are known to have. 

By considering several choices of initial KS states when the initial
interacting state is excited, we explored the effect of ISD on
dynamics in several exactly-solvable model systems and the performance
of the adiabatic approximation. We noted there are three sources of
error: excited KS initial states do not have the same density as the corresponding true excited states even when the exact functional is used, the use of the adiabatic
approximation to propagate the initial state, and the ground-state functional generating the adiabatic approximation itself being approximate. By
separating these errors from each other, we were able to properly
assign how badly they impacted the dynamics in model systems, allowing
us to see the effect of ISD more clearly.

When the initial KS state has a configuration that is significantly
different than the true state, the adiabatic approximation fails
severely -- even for non-interacting electrons, where there is a
significant error that can be interpreted as a static correlation
effect. The optimal choice for the KS initial state is one whose configuration is most similar to the true excited state; this is in fact the usual practise in recent calculations (e.g. Ref.~\cite{TRR05}). The error in using an adiabatic xc functional to propagate such a state, however, is significantly larger than the error the same functional produces when describing the propagation of an initial true ground-state, as demonstrated by the model soft-Coulomb helium atom example.
For two electrons, the spin-broken initial KS
state, with the two orbitals evolved under different spin-decomposed
potentials (depending on the instantaneous spin-densities), often
appeared to be the best choice, suggesting that orbital-specific
potentials (allowed by the generalized KS scheme) could be a useful
future avenue of research for TDDFT.   However, how to obtain such
potentials in a general $N$-electron case is not obvious (at least
without the need for empirical parameters).  Generally,
orbital-dependent functionals treated within the OEP may be promising
for these problems where memory-(of the density)-dependence, including
ISD at the KS level only, is naturally captured by the KS orbitals;
again, devising suitable orbital-dependent functionals with the
appropriate ISD is a direction for future research.

The example of coupled electron-nuclear dynamics in the model LiH
molecule, although supporting the earlier results of the optimal
choice of a KS initial state, above all, highlighted the need for more
accurate ground-state functional approximations for dissociation.

We do not provide in this work an approximation that includes ISD,
both of the true interacting state and the KS state; this is likely a
difficult task. But what is clear from the results in this paper, and
the stage of the field, is that now is the right time to address this
issue.

{\it Acknowledgments} Financial support from the National Science
Foundation (CHE-1152784), and a grant of computer time from the CUNY High
Performance Computing Center under NSF Grants CNS-0855217 and
CNS-0958379, are gratefully acknowledged.

\end{document}